\documentclass[a4paper,11pt]{article}
\usepackage{pos}
\usepackage{todonotes}
\usepackage{physics}
\usepackage{cleveref}

\title{Lattice artefacts on the Landau gauge gluon propagator from hypercubic tensor representations}
\ShortTitle{Lattice artefacts on the Landau gauge gluon propagator}

\renewcommand{\printHeadAuthors}{G. Catumba}
\author*[a,b]{Guilherme Catumba}
\author[b]{Orlando Oliveira}
\author[b]{Paulo J. Silva}

\affiliation[a]{Instituto de Física Corpuscular (IFIC) CSIC - Universitat de Valencia. \\ 46071, Valencia, Spain}

\affiliation[b]{CFisUC, Department of Physics, University of Coimbra \\ 3004-516 Coimbra, Portugal}

\emailAdd{gtelo@ific.uv.es}
\emailAdd{orlando@uc.pt}
\emailAdd{psilva@uc.pt}

\abstract{

Lattice tensor representations are used to investigate the lattice Landau gauge gluon propagator for the 4-dimensional pure SU(3) Yang-Mills gauge theory. 
Due to the different symmetry structure of hypercubic lattices compared to the continuum space-time, lattice correlation functions are described by different tensor structures. Therefore, form factors describing lattice correlation functions have, in principle, non-trivial relations with the continuum counterparts.
The use of several tensor bases respecting lattice symmetries, and the analysis of its completeness allows to quantify the deviations  of  the lattice results from the continuum theory, and also estimate the theoretical uncertainty in the propagator.  
Furthermore, our analysis tests continuum based relations with the lattice data and shows that the lattice Landau gauge gluon propagator is suitably described by a unique form factor, as in the continuum formulation.
Additionally, we identified classes of kinematic configurations where these deviations are minimal and the continuum description of lattice tensors is improved.
}

\FullConference{%
	The 38th International Symposium on Lattice Field Theory\\
	26th-30th July, 2021\\
	Zoom/Gather@Massachusetts Institute of Technology
}

\begin{document}
	\maketitle

\section{Introduction}

The gluon propagator is a gauge dependent Green function. Its study by various non-perturbative methods \cite{alex2020,papavassiliou2013effective,Orlando_spacing,leinweber,cornwall} suggests a dynamical generated mass that keeps the propagator finite in the Landau gauge.
From the point of view of lattice simulations, and to produce continuum results, a reliable description of the propagator requires a proper control of the discretization effects.

The most common approach to compute the lattice gluon correlator disregards the breaking of the O(4) symmetry on the lattice, and assumes that  its tensor structure is as in the continuum formulation, i.e.,
\begin{equation}
	\label{eq:continuum general basis}
	D_{\mu\nu}^{ab}(p) = \delta^{ab}D_{\mu\nu}(p) = \delta^{ab} \left(A(p^2)\delta_{\mu\nu} + B(p^2)\frac{p_\mu p_\nu}{p^2} \right)
\end{equation}
that in the Landau gauge reduces to
\begin{equation}
	\label{eq:continuum prop.}
	D_{\mu\nu}^{ab}(p) = \delta^{ab} \left( \delta_{\mu\nu} - \frac{p_\mu p_\nu}{p^2} \right)D(p^2).
\end{equation}
However, the analysis of lattice data for the propagator shows sizable discretization effects that can be described by the breaking of the O(4) group into one of its discrete subgroups, H(4), the symmetry group associated with an hypercubic four-dimensional lattice.

In order to reduce the discretization effects, it is common to replace the lattice momentum by an improved lattice momentum
\begin{equation}
	\label{eq:improved momentum}
	\hat p_\mu = \frac{2}{a}\sin\left( \frac{\pi n_\mu}{L_\mu} \right), ~n_\mu\in[ -L_\mu/2, L_\mu/2 ]
\end{equation}
(with $L_\mu$ the number of lattice points along $\mu$), that appear in the lattice perturbative solution for the gluon propagator. 
Moreover, an additional conical and cylindrical cut in momenta falling far away from the diagonal $(1,1,1,1)$ is considered, which further reduces the artefacts.

A second method uses the lattice momentum $p_\mu$ exploring the invariants of the H(4) group. While in the O(4) case there is a single linearly independent invariant, chosen as $p^2$, in the H(4) group, four independent invariants can be constructed:
\begin{equation}
	p^{[2]},~p^{[4]},~p^{[6]},~p^{[8]}, ~\text{with} ~p^{[n]} = \sum_{\mu}p_\mu^n.
\end{equation}
On the lattice, each scalar quantity $F(p^2)$ becomes a function of these, $F_{L}(p^{[2]},p^{[4]},p^{[6]},p^{[8]})$. 
In this work we disregard the $p^{[6]}$ and $p^{[8]}$ dependence and consider an extrapolation to $p^{[4]}=0$ with fixed $p^{[2]}$ in order to obtain the finite volume continuum limit $F(p^{[2]},0,0,0)$ up to $\order{a^2}$ \cite{fsoto2007}.
This requires various points with the same $p^2$, and so is applicable only in a limited range, that excludes the lower and higher momentum regions.

Usually one of these methods are used together with the continuum tensor structure of the propagator.	
However, the lattice gluon propagator is a second order symmetric tensor with respect to the H(4) group and not with respect to O(4).

Lattice tensor representations were previously used for the two and three point gluon correlation functions \cite{vujinov,vujimendes}, although focused on lower space-time dimensions.
Here we focus on applying the H(4) tensor bases to the description of the 4-dimensional lattice Landau gauge gluon propagator and study how the application of the improved momentum plus cuts, or the $p^{[4]}$ extrapolation  improves the description of the lattice propagator.
This approach allows to quantify the accuracy in the description of the lattice propagator by the continuum tensor when compared to more complete tensors.
The Landau gauge condition is also computed using the different tensor representations providing an additional test to the completeness.

\section{Lattice tensor representations for the gluon propagator}

The color space representation for the propagator  is  $D_{\mu\nu}^{ab}(p) = \delta^{ab}D_{\mu\nu}(p)$ and  still holds for the lattice formulation of QCD since $\delta^{ab}$ is the only symmetric second order SU(3) color tensor.
In order to build a second order symmetric tensor for the lattice it is necessary to first identify the lattice vectors, i.e. vectors with respect to H(4) transformations, and then the
second order tensors.

The elements of the H(4) group are rotations of $\pi/2$ around the axes of the hypercube, and also the inversion operations.
The vector-like elements under this symmetry group are the momentum $p_\mu$ or its odd-order powers. 
The improved momentum introduced before is exactly an infinite sum of odd powers of $ap_\mu/2$, thus being a proper hypercubic vector -- see \cite{vujinov,teloh4}.

Our interest lies in the symmetric second order tensors depending on a single momentum scale, $p$, which amounts to 10 linearly independent terms for the construction of a complete basis.
However, as will be evident, for the current statistics we must restrict to a smaller number of terms.
Hence, following \cite{vujinov}, a possible minimal (non-complete) lattice basis to describe the gluon propagator is
\begin{align}
	&D_{\mu\mu}(p) = J(p^2)\delta_{\mu\mu} + K(p^2)p_\mu^2, ~(\text{no sum}) \nonumber \\
	&D_{\mu\nu}(p) = L(p^2)p_\mu p_\nu,~\mu\neq\nu.
	\label{eq:partial_lattice_basis}
\end{align}
A possible extension of this basis considers additional form factors and higher order terms,
\begin{align}
	&D_{\mu\mu}(p) = E(p^2)\delta_{\mu\mu} + F(p^2)p_\mu^2 + G(p^2)p_\mu^4, ~(\text{no sum}) \nonumber \\
	&D_{\mu\nu}(p) = H(p^2)p_\mu p_\nu + I(p^2)p_\mu p_\nu(p_\mu^2 + p_\nu^2),~\mu\neq\nu ~(\text{no sum}).
	\label{eq:full_lattice_basis}
\end{align}
In both cases, the extraction of the form factors requires linear combinations of the propagator multiplied by various orders of the H(4) invariants -- this is done in detail in \cite{catumba2021gluon}.
The bases also do not restrict the form of momentum, and both forms of momentum, $p_\mu$ and $\hat p_\mu$, are considered.

Note that these form factors should be taken, in general, as functions of all H(4) invariants.
Further, although the form factors are gauge dependent, the form of the bases is gauge independent, with the gauge condition implying relations among the various functions.

To compare the faithfulness of each tensor bases, in addition to the direct comparison of the various form factors, a reconstruction procedure is used. The deviations to a complete description are probed by the ratio
\begin{equation}
	\mathcal{R} = \frac{\sum_{\mu\nu}|D^\text{\tiny orig}_{\mu\nu}|}{\sum_{\mu\nu}|D^\text{\tiny rec}_{\mu\nu}|},
	\label{eq:ratio}
\end{equation}
where $D^\text{\tiny orig}$ is the lattice propagator from the simulation, while $D^\text{\tiny rec}$ the reconstructed form factor, i.e., after projecting for a specific basis.
 
\section{Results for the Landau gauge gluon propagator}

\subsection{Extracting the propagator and lattice simulations}

The lattice data for the Landau gauge gluon propagator was generated with the Wilson action, at $ \beta = 6.0$, for a Monte Carlo simulation performed on a $80^4$ lattice with 550 gauge configurations. For this simulation the lattice spacing, measured from the string tension \cite{Bali1993}, is $a$ = 0.1016(25) fm or $1/a$ = 1.943(47) GeV. In all cases but the H(4) extrapolation, statistical errors are computed with the bootstrap method with a $67.5\%$ confidence level. For the $p^{[4]}$ extrapolation, the errors and the numbers reported are those obtained in a linear regression.

Since in the lattice formulation the fundamental variables are the group-valued links $U_\mu$, a projection to the algebra elements is required for the computation of the gluon propagator.
After gauge fixing to the Landau gauge \cite{teloh4} the gluon field is computed by
\begin{equation}
	A_\mu(x+a\hat\mu/2) = \eval{ \frac{U_\mu(x) - U_\mu^\dagger(x)}{2iag} }_\text{traceless}.
\end{equation}
Further, the momentum space field is obtained by Fourier transform, and the lattice gluon propagator is obtained as a Monte-Carlo ensemble average
\begin{equation}
	D_{\mu\nu}^{ab}(p) =  \frac{1}{V}\expval{A_\mu^a(p) A_\nu^b(p')} \delta(p+p').
\end{equation}
Further details of the simulation are given in \cite{teloh4,catumba2021gluon}.

\subsection{Continuum description and general correction methods}

\begin{figure}[t]
	\centering
	\includegraphics[scale=0.26]{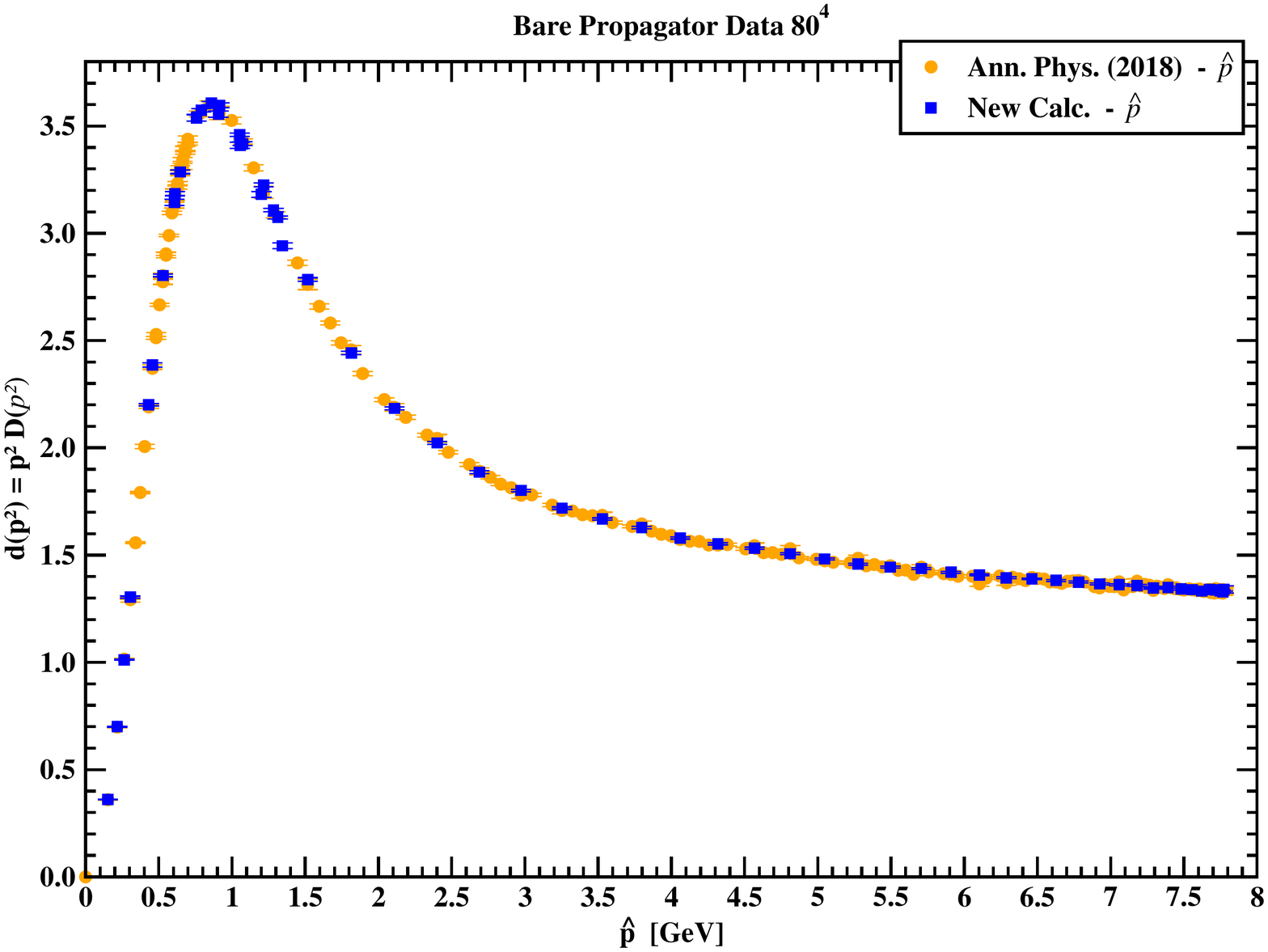}
	\includegraphics[scale=0.26]{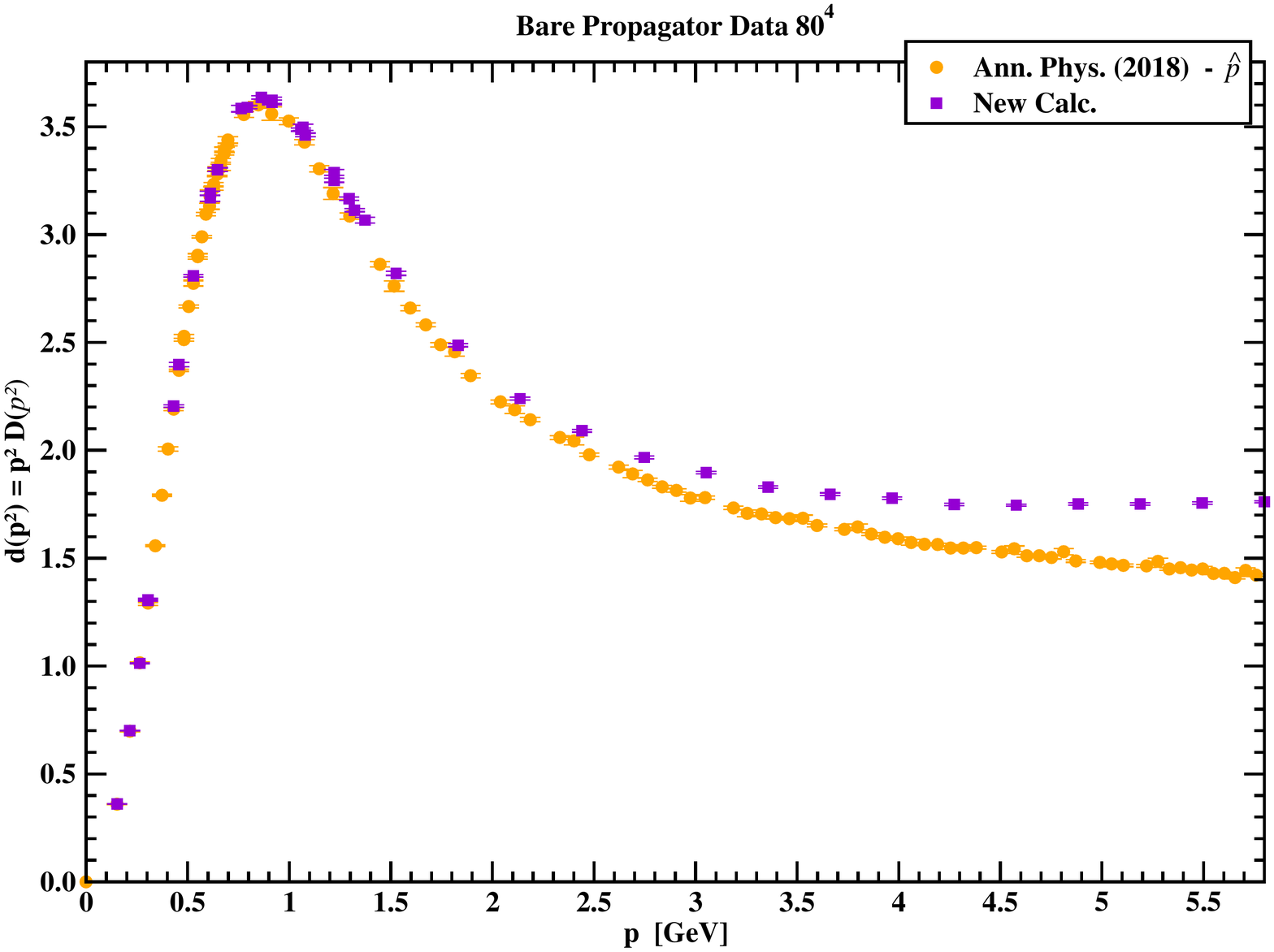} 
	\includegraphics[scale=0.26]{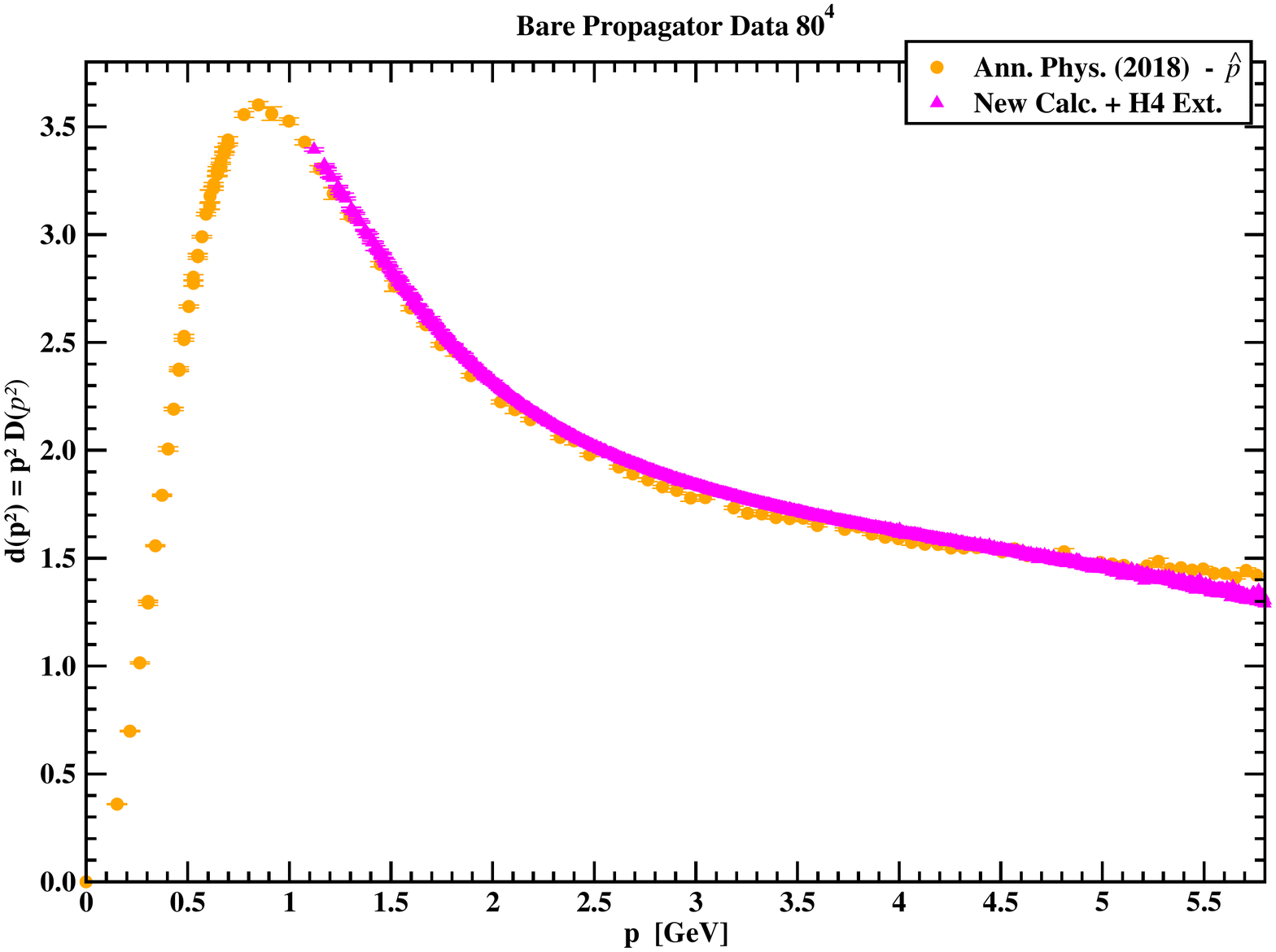} 
	\caption{Landau gluon dressing function $d(p^2) = p^2 D(p^2)$ using the continuum tensor basis and after momentum cuts for the improved momentum (top left), for the lattice momentum (top right), and after performing the H4 extrapolation with lattice momentum (bottom). All the results are compared to reference data (yellow) \cite{Dudal_2018}.}
	\label{fig:correction methods}
\end{figure}

We begin by illustrating the general correction methods to the discretization effects. This is done for the continuum tensor in \cref{eq:continuum prop.} but the conclusions remain valid for any other tensor structure. The comparison of the methods is shown in \cref{fig:correction methods}, where the momentum cuts with both forms of momenta and also the $p^{[4]}$ extrapolation are compared with the reference data from \cite{Dudal_2018}, always shown as a function of $\hat p$ and after conical and cylindrical momentum cuts.
%
%
The new computation uses an average over all lattice points with the same H(4) invariants, producing smaller statistical errors, when compared with the previous reference data.
This shows the relevance of using the remaining symmetry to improve the signal to noise ratio on the correlation functions.
It is also clear the difference in the shape of the form factor when using the improved lattice momentum.
Note that the lattice data together with the $p^{[4]}$ extrapolation also reproduces the reference data with good statistical precision, although for a smaller range.

\subsection{Lattice form factors}

\begin{figure}[t]
	\centering
	\includegraphics[scale=0.26]{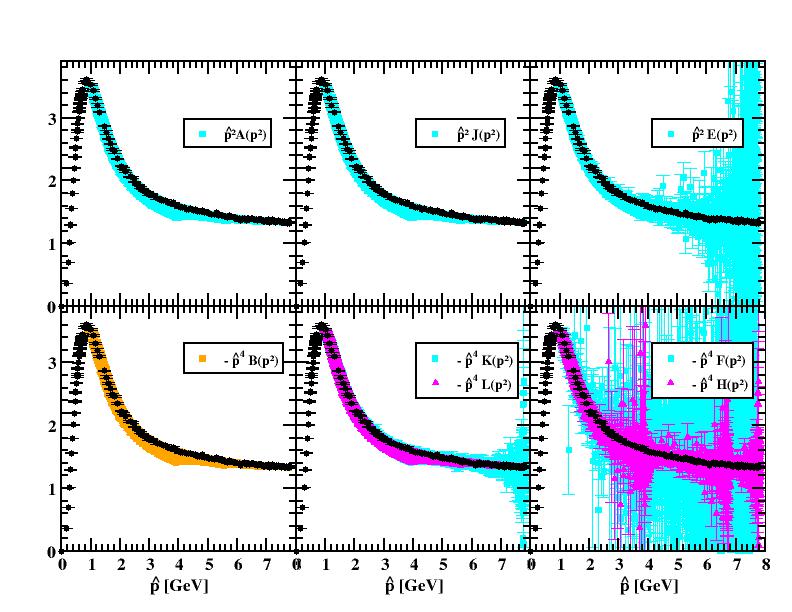} 
	\includegraphics[scale=0.26]{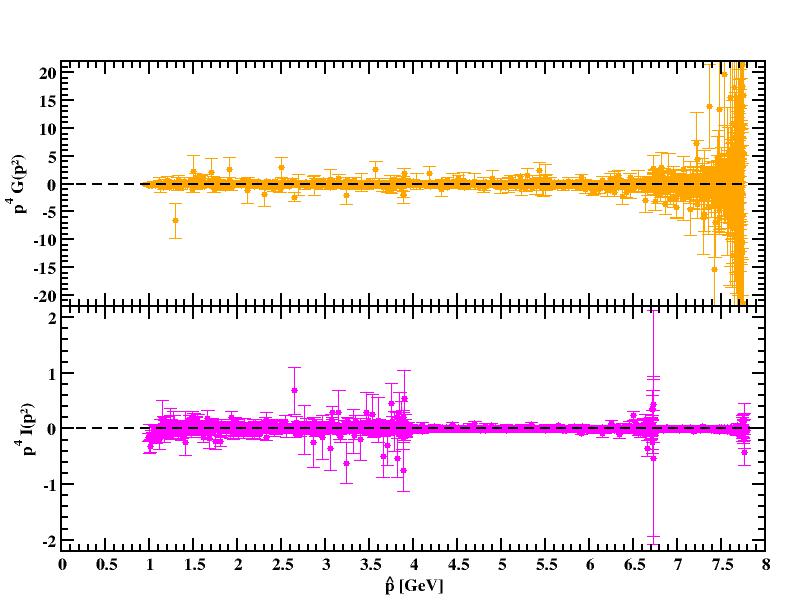} 
	\caption{Form factors from the continuum, \cref{eq:continuum general basis} and lattice tensor bases, \cref{eq:partial_lattice_basis,eq:full_lattice_basis} using improved momentum without cuts along the diagonal. In black is shown the reference data from \cite{Dudal_2018}.}
	\label{fig:formfact_pimp}
\end{figure}

\begin{figure}[t]
	\centering
	\includegraphics[scale=0.26]{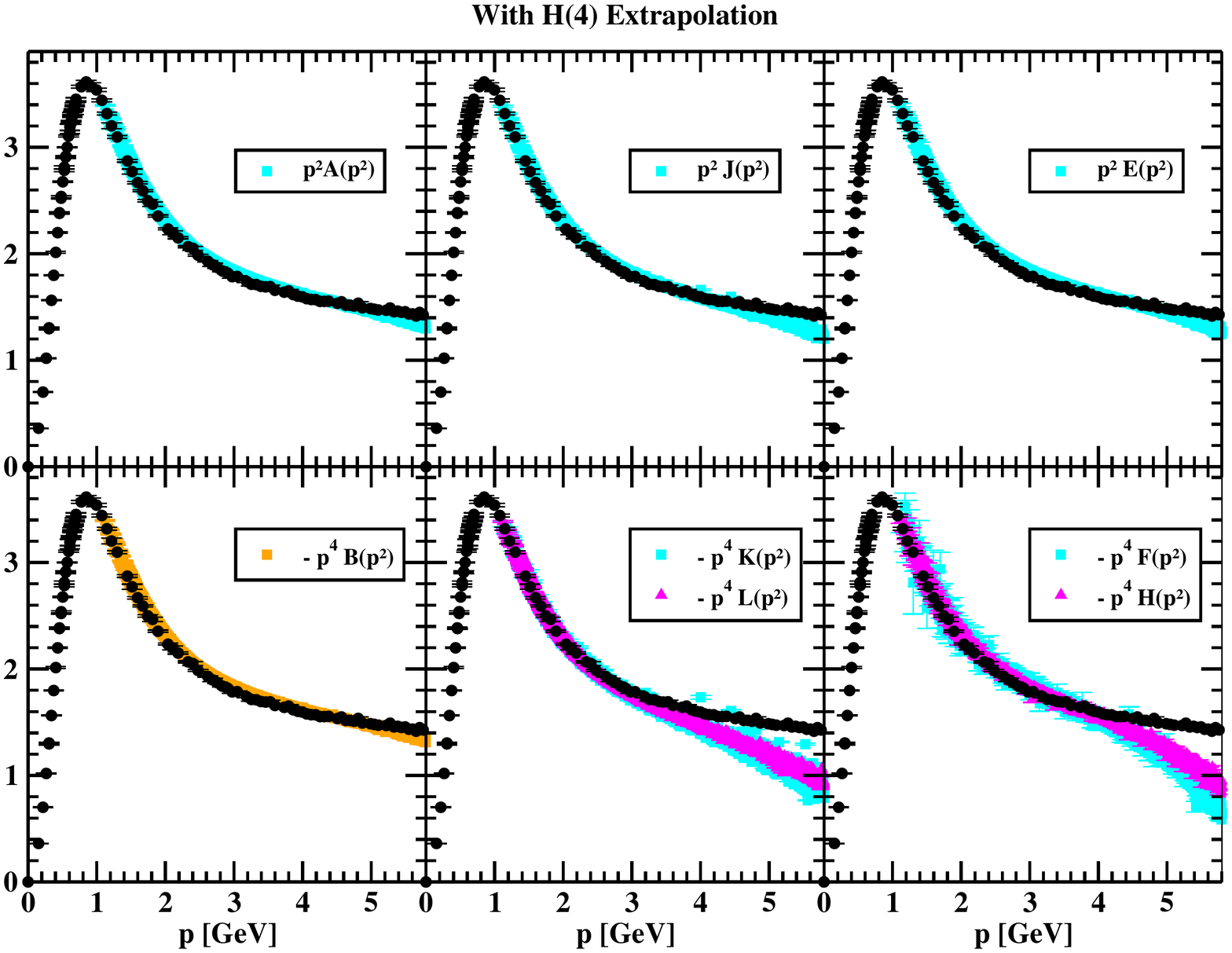} 
	\includegraphics[scale=0.26]{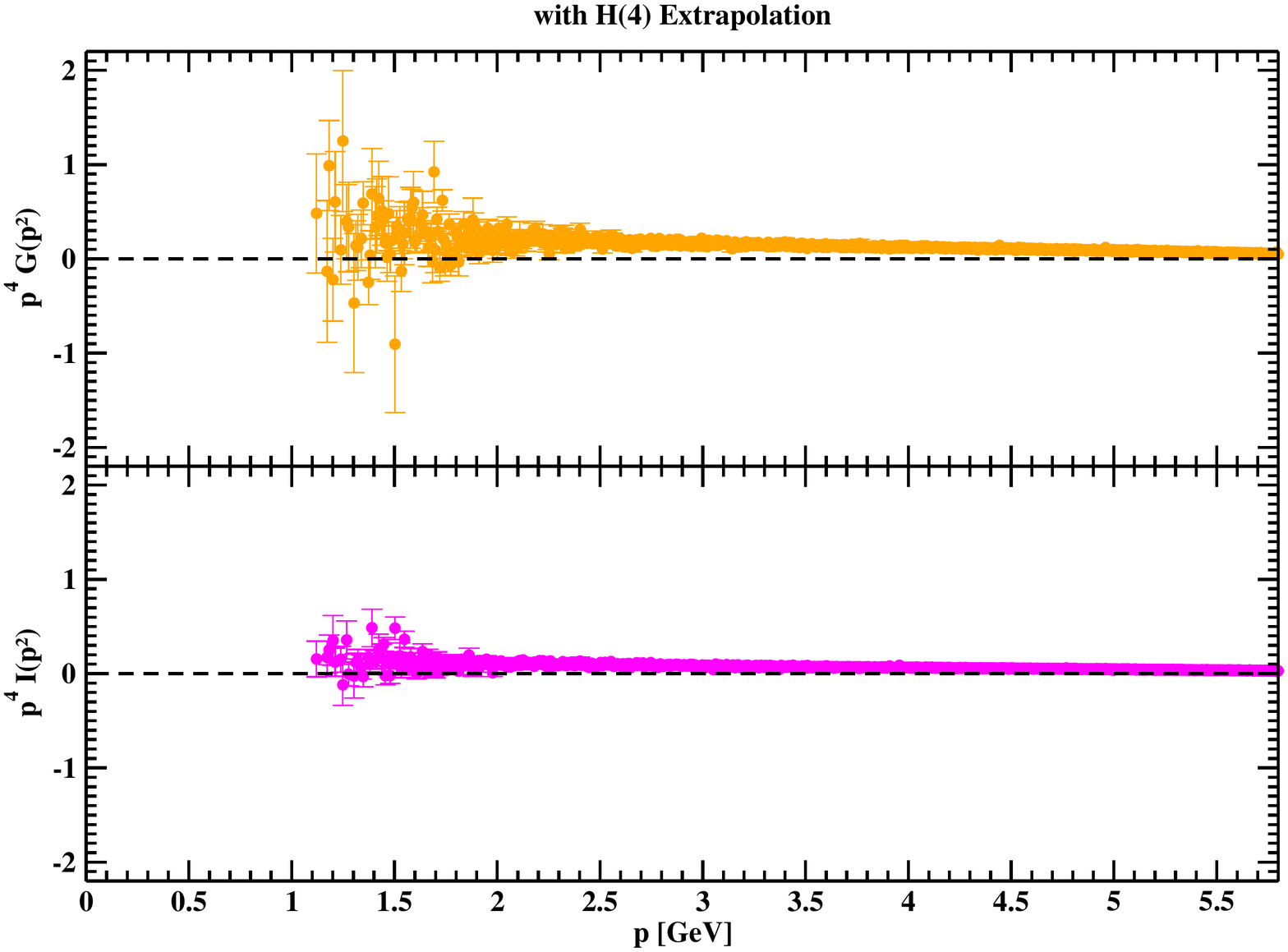} 
	\caption{Same as in \cref{fig:formfact_pimp} with lattice momenta after the H4 extrapolation.}
	\label{fig:formfact_h4ext}
\end{figure}

These two correction methods are also applicable to other, more complete tensor bases, such as \cref{eq:continuum general basis,eq:partial_lattice_basis,eq:full_lattice_basis}.
The results are shown in \cref{fig:formfact_pimp,fig:formfact_h4ext} with the cuts applied to the formulations using improved momentum, and the H(4) extrapolation applied to the lattice momentum case, respectively.
Although momentum cuts were also applied to the non-continuum tensor bases, its effect is similar to the one in \cref{fig:correction methods} however more affected by statistical fluctuations.
%
Note that close to the continuum, the several form factors from the various bases should satisfy 
\begin{align}
	&E(p^2) \rightarrow D(p^2), && J(p^2) \rightarrow D(p^2), \\
	&-p^2F(p^2), ~ -p^2H(p^2) \rightarrow D(p^2), && -p^2K(p^2), ~ -p^2L(p^2) \rightarrow D(p^2),\\
	&G(p^2), ~ I(p^2) \rightarrow 0.
	\label{eq:continuum relations extended}
\end{align}
For the higher order form factors, $p^4I(p^2)$ and $p^4G(p^2)$, the continuum relations are obtained for the full range of momentum for the improved momentum case only.
In what concerns the H(4) extrapolation, these relations are only satisfied for lower momenta.
Additionally, the error bars in $G(p^2)$ are larger than for $I(p^2)$ due to the mixing with other form factors in the projection.

The remaining form factors are shown in the left plots of \cref{fig:formfact_pimp,fig:formfact_h4ext} in a way to test deviations from the continuum relations.
For the improved momentum formulation and for higher momenta, the noise affecting the form factors in the larger basis do not allow to disentangle the lattice artifacts. 
The use of lattice momentum with the extrapolation seems to substantially reduce the noise while satisfying the continuum relations. However, again these are limited to a finite range of momentum since the extrapolation clearly fails to deliver reliable results for $p \gtrsim 5$ GeV.
Furthermore, the increase in the number of form factors carries an increase in the noise affecting the data, which prevented us from using larger tensor bases.

\subsection{Completeness and orthogonality}

\begin{figure}[t]
	\centering
	\includegraphics[scale=0.26]{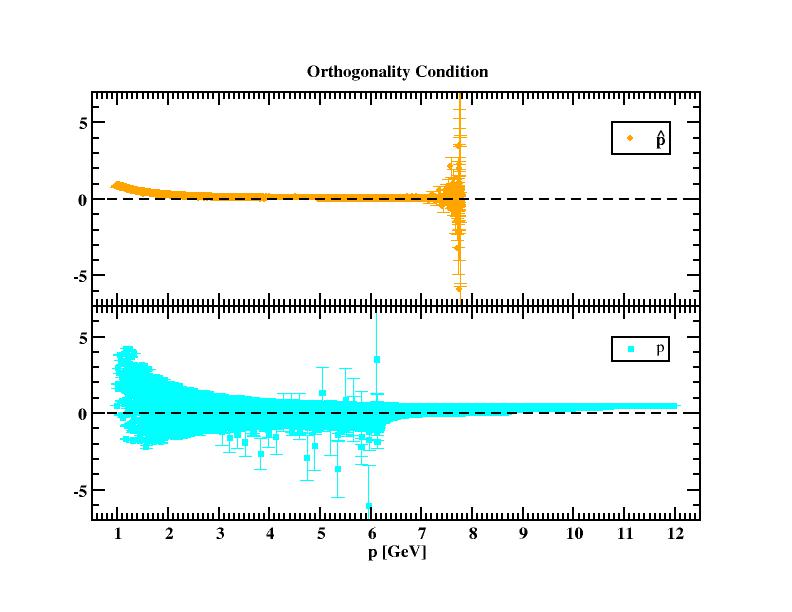}
	\raisebox{0.06\height}{\includegraphics[scale=0.23]{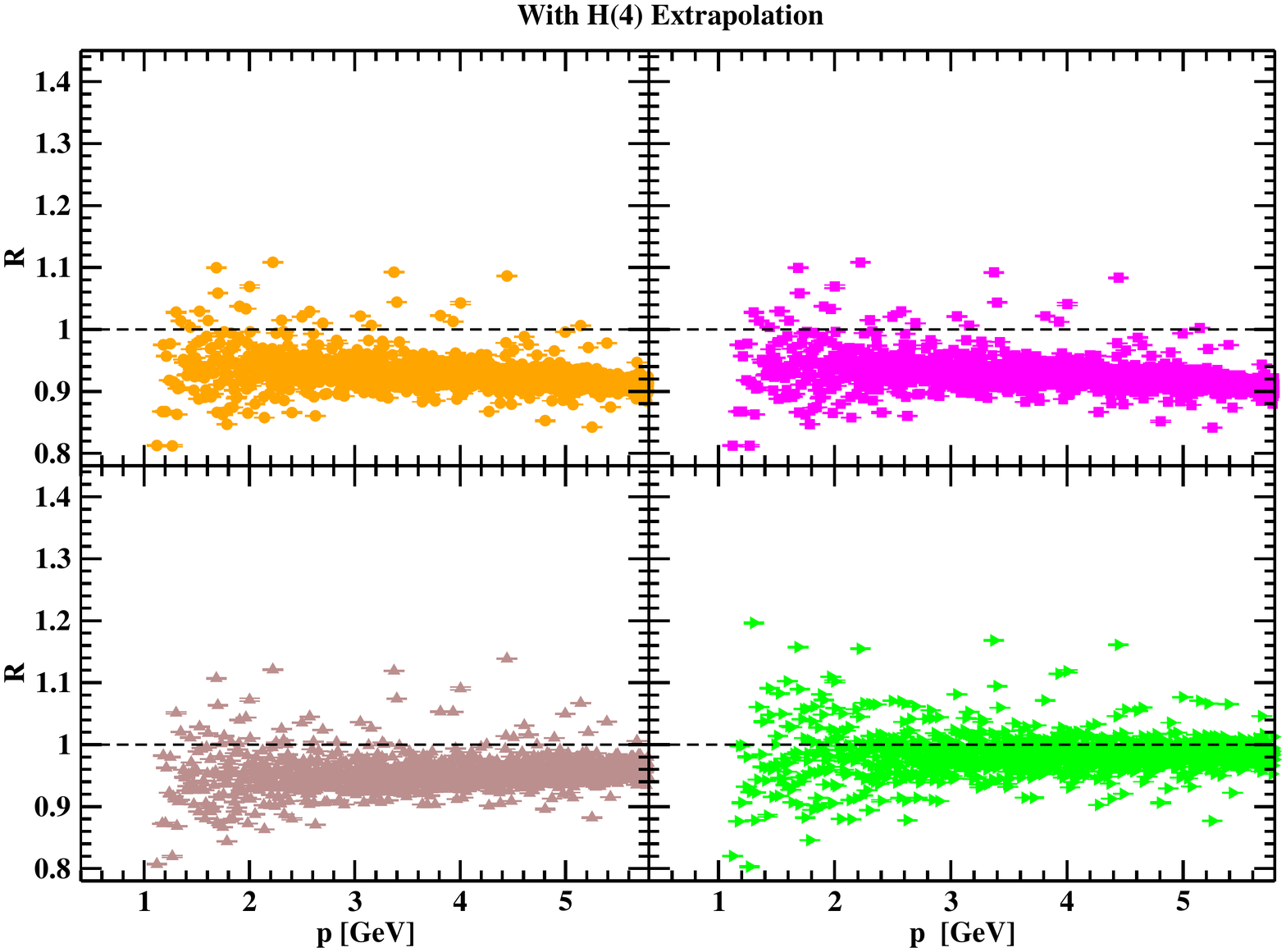}}
	\caption{Left plot: Orthogonality condition, $p_\mu D_{\mu\nu}(p)$ for the improved (top) and lattice momentum (bottom); Right plot: $\mathcal{R}$ for the single form factor continuum basis \cref{eq:continuum prop.} (top left), for the general continuum basis \cref{eq:continuum general basis} (top right), and the minimal and extended lattice tensor bases \cref{eq:partial_lattice_basis,eq:full_lattice_basis} (bottom left and right, respectively).}
	\label{fig:orthogonality}
\end{figure}

In the Landau gauge, the gluon field is orthogonal to its momentum, $p_\mu A_\mu(p) = 0$. This constrains the relations between the form factors, which for the largest basis and $p_\nu \neq 0$ reads
\begin{equation}
	\sum_{\mu}p_\mu D_{\mu\nu}(p)= E + p_\nu^2F + p_\nu^4G + (p^2-p_\nu^2)H + \left(p^{[4]} + p^2p_\nu^2 - 2p_\nu^4\right)I = 0.
	\label{eq:orthogonality_general}
\end{equation}
This condition is shown in the left-hand side of \cref{fig:orthogonality} for both types of momentum. While it seems to be better fulfilled with the improved momentum, we see that the matching is not exact, with the points departing from zero for lower momentum.
%
The completeness of each basis reports on the faithfulness in reproducing the correct lattice tensor.
In \cref{fig:orthogonality}, the complete $\mathcal{R}$ data after the H(4) extrapolation as a function of lattice momentum is shown. There is an increase in the completeness, i.e. values of $\mathcal{R}$ closer to 1, for larger lattice bases. 
Notice however, that even for the smallest tensor basis, the completeness fluctuates around $0.9$.

\begin{figure}[t]
	\centering
	\includegraphics[scale=0.26]{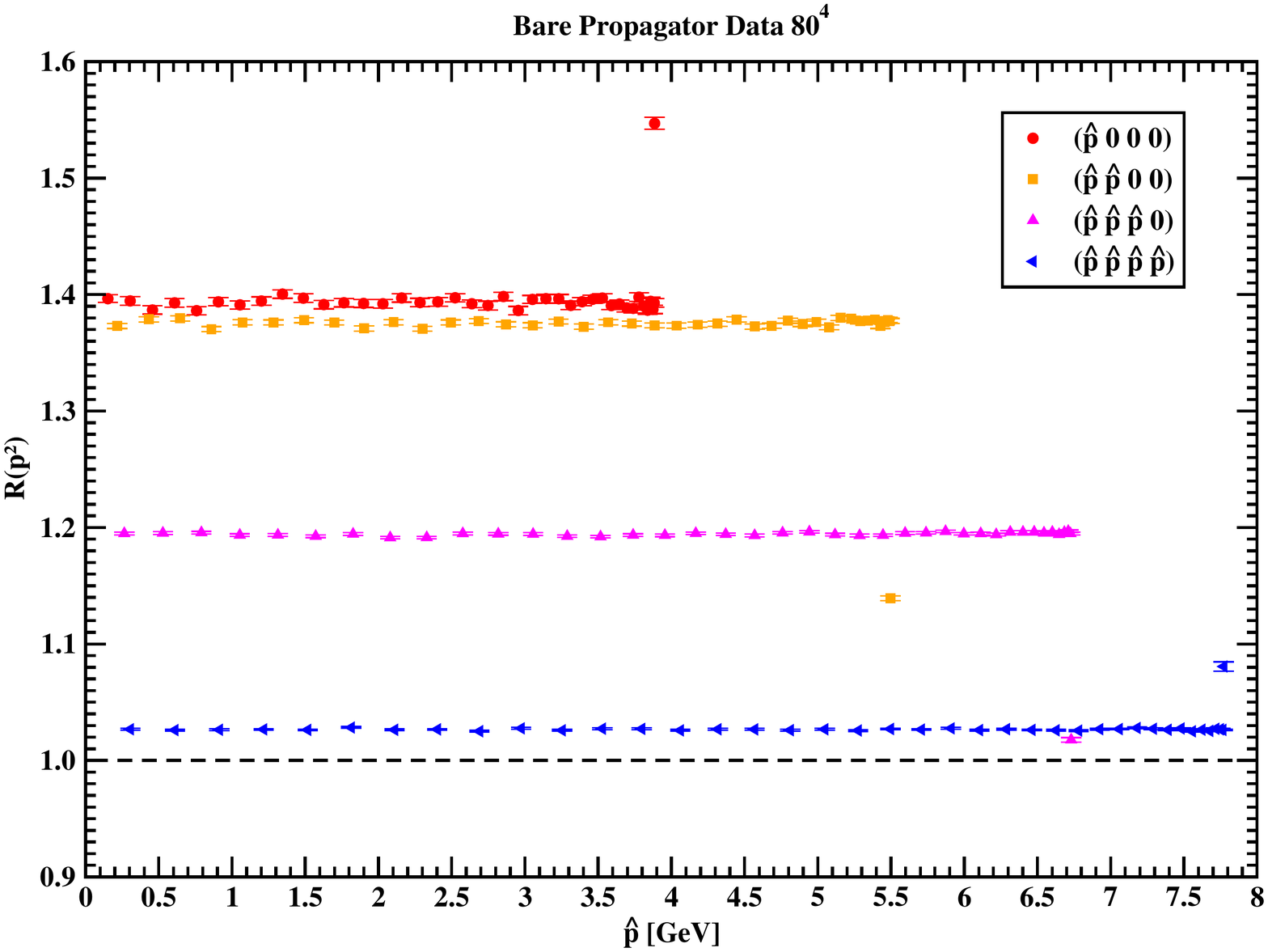} 
	\includegraphics[scale=0.26]{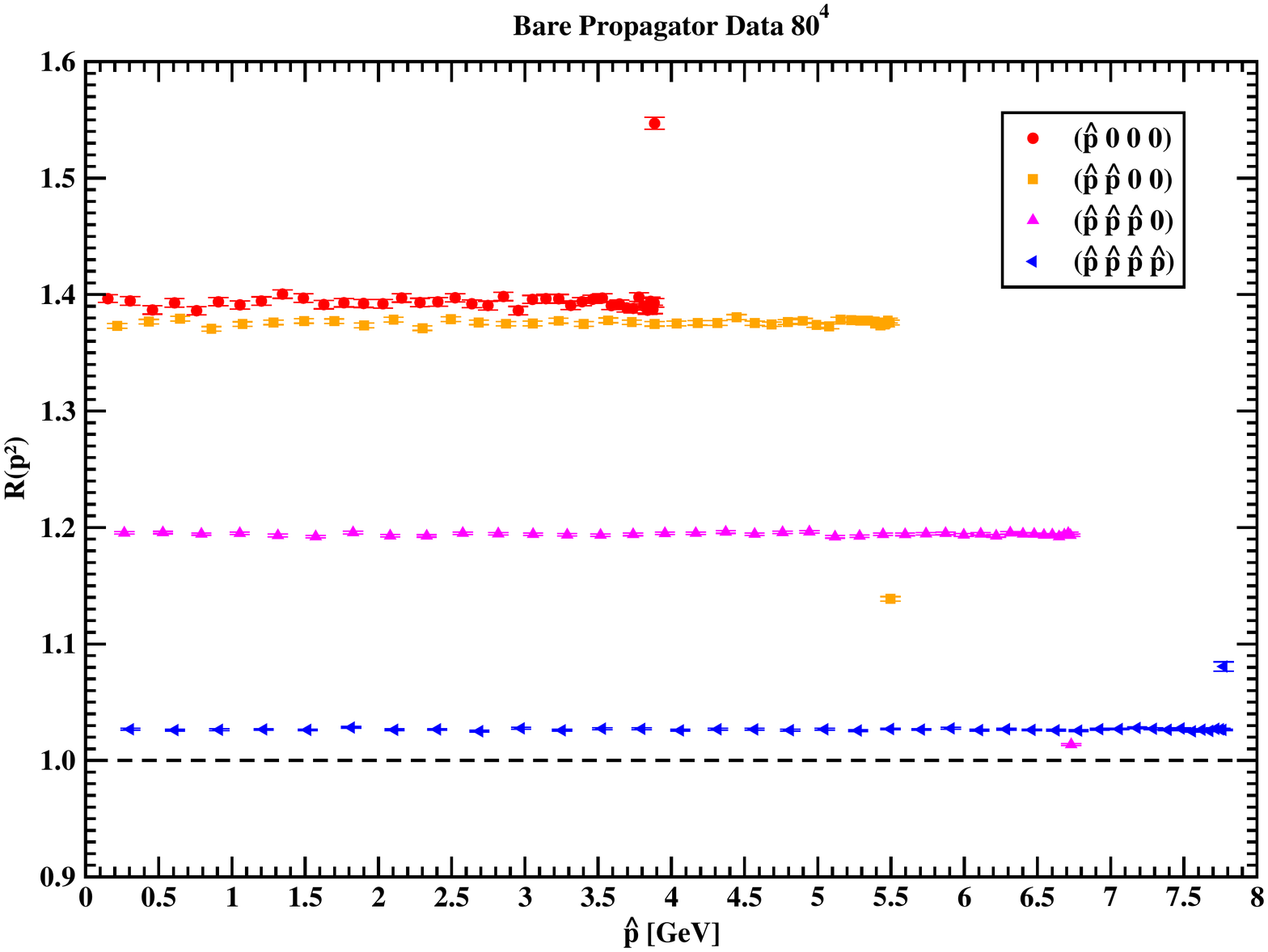} 
	\caption{Reconstruction ratio for different classes of single scale improved momenta  for the continuum basis (left) and for the extended basis (right).}
	\label{fig:completeness diag}
\end{figure}

By analyzing the completeness of each basis we notice that there are special momentum configurations where the continuum relations are better fulfilled. 
Shown in \cref{fig:completeness diag} are the results of the completeness for single scale type of momentum points. We see that the use of lattice or improved momentum is irrelevant in this case, with both the smaller and larger bases providing the same $\mathcal{R}$. Additionally, this seems to depend solely on the distance of the momentum from the diagonal of the lattice, thus corroborating the usual method of conical and cylindrical cuts.

\section{Conclusion}

In this work we explore the rotational symmetry breaking of the lattice formulation of QCD in the construction of tensor representations for the Landau gauge gluon propagator. 
In general, the use of the H(4) group symmetry in the computation of the correlator improves the Monte-Carlo signal to noise ratio.


Continuum-like relations between the bases form factors in \cref{fig:formfact_pimp,fig:formfact_h4ext} are satisfied within error bars.
All form factors are either compatible and reproduce the reference data, or vanish for most of the momentum range.
The noise affecting these functions is larger for the extended basis, with the H4 extrapolation producing a better signal.

The ratio $\mathcal{R}$ allows to evaluate how well each tensor basis describes the lattice propagator.
As seen in  \cref{fig:orthogonality}, the use of a larger tensor produces a better characterization of the propagator. 
However, for the current statistical precision, completeness is never fully achieved due to the noise level.
The overall analysis to the form factors, completeness, and orthogonality, shows that the usual approach, based on the use of the continuum basis with additional improved momentum cuts, provides a reliable estimation of the continuum limit of the gluon propagator $D(p^2)$.


\acknowledgments
	
This work was supported by national funds from FCT – Fundação para a Ciência e a Tecnologia, I.P., within the
projects UIDB/04564/2020 and UIDP/04564/2020. G.T.R.C. acknowledges financial support from FCT (Portugal)
under the project UIDB/04564/2020. P. J. S. acknowledges financial support from FCT (Portugal) under Contract
No. CEECIND/00488/2017.
This work was granted access to the HPC resources of the PDC Center for High Performance Computing at the KTH Royal Institute of Technology, Sweden, made available within the Distributed European Computing Initiative by the PRACE-2IP, receiving funding from the European Community’s Seventh Framework Programme (FP7/2007-2013) under grand agreement no. RI-283493. The use of Lindgren has been provided under DECI-9 project COIMBRALATT.
We acknowledge that the results of this research have been achieved using the PRACE-3IP project (FP7 RI312763) resource Sisu based in Finland at CSC. The use of Sisu has been provided under DECI-12 project COIMBRALATT2.
We also acknowledge the Laboratory for Advanced Computing at the University of Coimbra (http://www.uc.pt/lca) for providing access to the HPC resource Navigator.
In addition, G.T.R.C. acknowledges financial support from the Generalitat Valenciana (genT program CIDEGENT/2019/040) and Ministerio de Ciencia e Innovacion  PID2020-113644GB-I00.

\end{document}